\documentclass[11pt]{article}
\usepackage{moriond,epsfig}

\def\Journal#1#2#3#4{{#1} {\bf #2}, #3 (#4)}

\def\PRL{\em Phys. Rev. Lett.}
\def\PRD{{\em Phys. Rev.} D}

\def\JHEP{{\em JHEP}}

\def\be{\begin{equation}}
\def\ee{\end{equation}}
\def\bea{\begin{eqnarray}}
\def\eea{\end{eqnarray}}

\newcommand{\Zmumu}   {\mbox{${\mathrm Z}(\rightarrow\mu \mu$)+jets}}

\newcommand{\Zll}   {\mbox{${\mathrm Z}(\rightarrow l l$)}}
\newcommand{\ttbar}   {\mbox{${t\bar{t}}$}}

\newcommand{\Znunu}   {\mbox{${\mathrm Z}(\rightarrow\nu \nu$)}}

\newcommand{\Wlnu}   {\mbox{${\mathrm W}(\rightarrow l\nu$)}}
\newcommand{\Wmunu}   {\mbox{${\mathrm W}(\rightarrow \mu\nu$)}}

\newcommand{\Wenu}   {\mbox{${\mathrm W}(\rightarrow e\nu$)}}
\newcommand{\pt}{\ensuremath{\mathrm{p_T}}}
\newcommand{\Met}{\mbox{$\not\!\! E_{\mathrm{T}}$}}
\begin{document}
\vspace*{4cm}
\title{Search for dark matter at the LHC using missing transverse energy}

\author{Sarah Alam Malik }

\address{1230 York Avenue  New York, NY 10065, USA}

\maketitle\abstracts{
Results are presented for a search for dark matter at the LHC using the signatures of a monojet plus missing transverse energy and a monophoton plus missing transverse energy. The data were collected by the CMS detector at the LHC with pp collisions at a centre-of-mass energy of 7 TeV and an integrated luminosity of 5 fb$^{-1}$. In the absence of an excess of events in the data compared to the Standard Model prediction, limits are set on the dark matter-nucleon scattering cross section which can be directly compared with bounds from the direct detection experiments. } 

\section{Introduction}
There is strong evidence from numerous cosmological observations, such as the rotational speed of galaxies and gravitational lensing, that approximately 25$\%$ of the matter-energy density of the universe is made up of dark matter. 

 One of the most popular class of candidates for dark matter are Weakly Interacting Massive Particles (WIMPs). Pairs of these particles can be produced at the LHC if the partonic centre-of-mass energy is above the energy threshold for a WIMP pair. When accompanied by the emission of a jet or a photon from the initial state, these processes lead to the signature of a jet plus missing transverse energy~\cite{bib:monojet} and a photon plus missing transverse energy~\cite{bib:monophoton}. This article will describe searches for dark matter performed at the Large Hadron Collider (LHC) using data collected by the Compact Muon Solenoid (CMS) detector corresponding to an integrated luminosity of 5 fb$^{-1}$. 

Searches for dark matter at colliders have previously been performed in the context of new physics models such as Supersymmetry (SUSY) where the lightest SUSY particle is neutral, stable, and thus a good candidate for dark matter. Recently, more model-independent approaches have been considered in which no assumptions are made about the new physics model that is producing the WIMPs. This is achieved by working within the framework of an effective field theory and assuming that the dark matter particle is the only new state that is accessible to the LHC. The mediator that couples the dark matter particles to the Standard Model (SM) states is assumed to be very heavy such that it can be integrated out and the interaction can be treated as a contact interaction. This approach is described in~\cite{bib:tait,bib:tait2,bib:tait3,bib:tait4,bib:roniTevatron,bib:roniLHC}.

Working with one of these models~\cite{bib:roniTevatron,bib:roniLHC}, we assume that the dark matter particle($\chi$) is a Dirac fermion and the contact interaction is characterized by a scale $\Lambda = M/\sqrt{g_{\chi}g_{{q}}}$ where $M$ is the mass of the mediator and $g_{\chi}$ and $g_{{q}}$ are its couplings to $\chi$ and quarks, respectively. Operators that describe the nature of the coupling of the mediator can be defined and two possibilities are considered~\cite{bib:roniTevatron,bib:roniLHC}, a vector operator and an axial-vector operator, which proceed via an s-channel exchange of a new heavy gauge boson and translate to bounds on the spin independent and spin dependent dark matter-nucleon interactions, respectively.

\section{Monojet search}
The data used in the study of events with a monojet and missing transverse energy is recorded by a trigger that requires $\Met > 80 (95)$ and atleast one jet with $p_{T} > 80$ GeV. The trigger is found to be fully efficient for $\Met > 200$ GeV and jet $\pt > 110$ GeV. Particles in an event are individually identified as either charged hadrons, neutral hadrons, photons, muons, or electrons, using a particle-flow reconstruction~\cite{bib:particleflow}. The reconstructed particles are fully calibrated and clustered into jets using the anti-$k_{T}$ algorithm~\cite{bib:antikt} with a distance parameter of 0.5. The \Met\ in this analysis is defined as the magnitude of the vector sum of all particles reconstructed in the event excluding muons. This definition allows the use of a control sample of \Zmumu\ events to estimate the \Znunu\ background.
 A Monte Carlo study of the optimization of the \Met\ cut found the optimal cut producing the best limits on the dark matter signal to be $\Met > 350$ GeV. 
The signal sample is therefore selected by requiring $\Met > 350$ GeV and the jet with the highest transverse momentum to have $\pt > 110$ GeV and $|\eta| < 2.4$. A second jet is allowed, as signal events typically contain an initial or final state radiated jet, provided its angular separation in azimuth from the highest $\pt$ jet satisfies $\Delta\phi < 2.5$ radians. This angular requirement suppresses QCD dijet events. Events containing more than two jets with $\pt > 30$ GeV are vetoed. This requirement suppresses \ttbar\ and QCD multijet events. To reduce background from electroweak processes and top-quark decays, events with isolated muons and electrons with $\pt > 10$ GeV are rejected. Events with an isolated track with $\pt > 10$ GeV are also removed as they come primarily from the decay of $\tau$-leptons. A track is considered isolated if the scalar sum of the transverse momentum of all tracks with $\pt > 1$ GeV in the annulus of radius $0.02 < \Delta R < 0.3$ around its direction is less than 1$\%$ of its $\pt$.


The dominant backgrounds to this search are from Z+jet and W+jet events, where the Z boson decays to a pair of neutrinos and the W decays leptonically.
These backgrounds are estimated from a control sample of $\mu$+jet events, where \Zmumu\ events are used to estimate \Znunu\ and \Wmunu\ events are used to estimate the remaining W+jets background. 
The \Znunu\ background is estimated by correcting the observed \Zmumu\ event yield by the detector acceptance, the selection efficiency and the ratio of the branching fractions. It is estimated to be 900 $\pm$ 94 events, where the uncertainty includes statistical and systematic contributions. The dominant uncertainty is from the statistical size of the \Zmumu\ control sample. 
The remaining W+jets background is estimated by selecting a control sample of \Wmunu\ events and correcting for the inefficiencies of the electron and muon selection requirements. The W+jets background is estimated to be 312 $\pm$ 35 events.

Background contributions from QCD multijet events, \ttbar\ and \Zll+jets production are small and are obtained from the simulation. A 100$\%$ uncertainty is assigned to these backgrounds. 

Table~\ref{tab:monojet_results} shows the estimated contributions to the monojet sample from the SM backgrounds and the total number of observed events. The observed event yield is consistent with the number of events expected from SM backgrounds. 
\begin{table}[t]
\caption{Event yields for SM background predictions and the observed number of monojet events in the data.\label{tab:monojet_results}}
\vspace{0.4cm}
\begin{center}
\begin{tabular}{|c|c|}
\hline
Source & Estimate\\ \hline
\Znunu+jets & 900 $\pm$ 94  \\
W+jets & 312 $\pm$ 35 \\
\ttbar & 8 $\pm$ 8 \\
\Zll+jets & 2 $\pm$ 2 \\
Single top & 1 $\pm$ 1 \\
QCD Multijets & 1 $\pm$ 1 \\ \hline
Total background & 1224 $\pm$ 101 \\
Observed candidates & 1142 \\ \hline
\end{tabular}
\end{center}
\end{table}

\section{Monophoton search}
The final state containing a photon and \Met\ is also a signature of many new physics scenarios, including the production of dark matter particles. 
The dataset is collected by single-photon triggers that are fully efficient for the selected signal region. A photon candidate is selected by requiring $\pt > 145$ GeV and $|\eta| < 1.44$, to ensure it is in the central barrel region of the detector where purity is highest. The ratio of the energy deposited in the hadronic calorimeter (HCAL) to that in the electromagnetic calorimeter (ECAL) within a cone of $\Delta R = 0.15$ is required to be less than 0.05, where $\Delta R = \sqrt{(\Delta \phi)^{2} + (\Delta \eta)^{2}}$ is defined relative to the photon candidate and the azimuthal angle $\phi$ is measured in the plane perpendicular to the beam axis. Photon candidates are also required to be isolated and have a shower distribution in the ECAL that is consistent with that expected for a photon.   
The \Met\ is defined as the magnitude of the vector sum of the transverse energies of all the reconstructed objects in the event, as computed using the particle-flow algorithm and is required to be $\Met > 130$ GeV. 
To reduce instrumental background arising from showers induced by muons in the beam halo or cosmic rays, the energy deposited in the crystal containing the largest signal within the photon is required to be within $\pm 3$ ns of the time expected for particles from a collision. Spurious signals in the ECAL are eliminated by requiring the energy deposition times for all crystals within an electromagnetic shower to be consistent. Events are also vetoed if they contain significant hadronic activity~\cite{bib:monophoton}.

Backgrounds processes contributing to the photon plus \Met\ topology include; $Z+\gamma$ production where the Z decays to neutrinos, $W+\gamma$ production where the W decays leptonically, \Wenu\ events where the electron is misidentified as a photon, diphoton production, QCD multijet events where one of the jets mimics a photon and another is mismeasured, and events from out of time collisions.

Backgrounds from out of time collisions, \Wenu\ events and QCD multijet events are estimated from the data. Other backgrounds from $\Znunu+\gamma$, $\Wlnu+\gamma$, $\gamma+$jet and diphoton events are estimated from simulation. Table~\ref{tab:monophoton_results} shows the estimated contributions to the signal sample from the various SM processes and the observed candidate events. The number of events observed in the data are consistent with the expected number of events from SM backgrounds. 
\begin{table}[t]
\caption{Event yields for SM background predictions and the observed number of monophoton events in the data.\label{tab:monophoton_results}}
\vspace{0.4cm}
\begin{center}
\begin{tabular}{|c|c|}
\hline
Source & Estimate \\ \hline
$\Znunu+\gamma$ & 45.3 $\pm$ 6.9 \\ 
Jet mimics photon  & 11.2 $\pm$ 2.8 \\
Beam halo & 11.1 $\pm$ 5.6 \\
Electron mimics photon & 3.5 $\pm$ 1.5 \\
$W\gamma$ & 3.0 $\pm$ 1.0 \\
$\gamma\gamma$ & 0.6 $\pm$ 0.3 \\
$\gamma$+jet & 0.5 $\pm$ 0.2 \\ \hline
Total background & 75.1 $\pm$ 9.5 \\
Observed candidates & 73 \\ \hline
\end{tabular}
\end{center}
\end{table}

\section{Interpretation}

\begin{figure}[t]
\begin{center}
\includegraphics[scale=0.38]{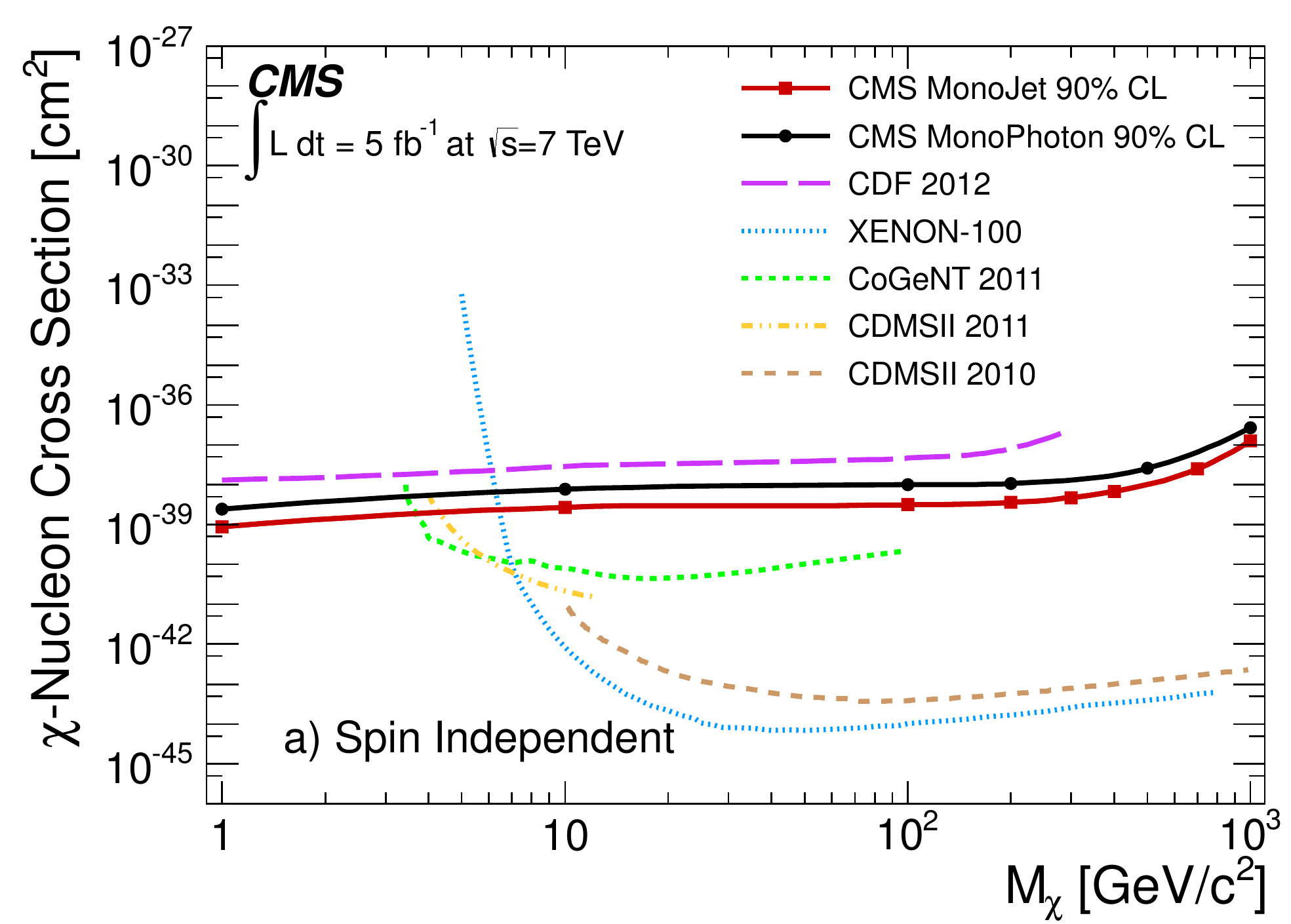}
\includegraphics[scale=0.38]{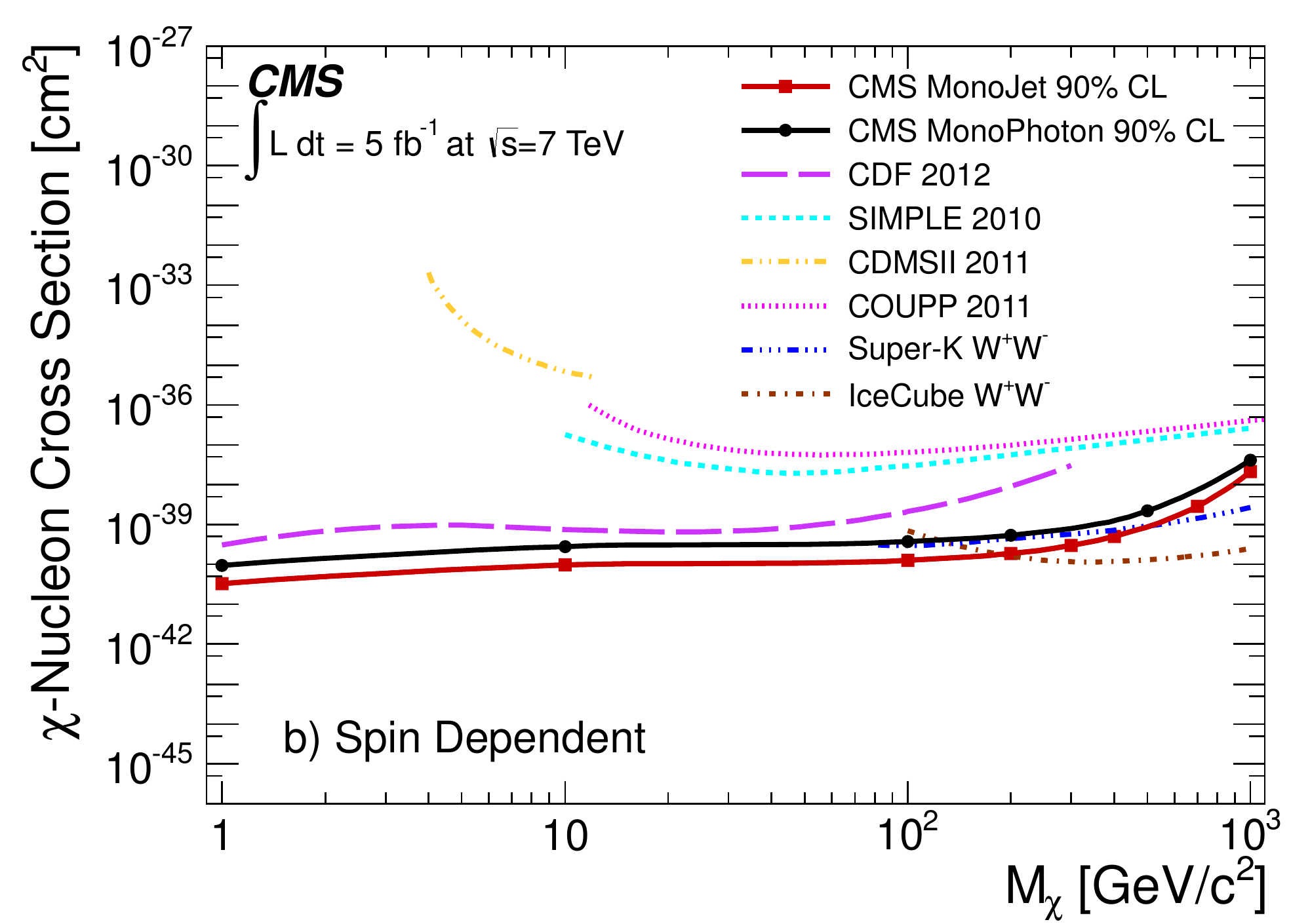}
\caption{The 90$\%$ CL upper limits on the dark matter-nucleon cross section as a function of $M_{\chi}$ for (a) spin independent and (b) spin dependent interactions. Also shown are the bounds from other experiments.}
\label{fig:limits}
\end{center}
\end{figure}

Since there is no observed excess of events in the data over those expected from SM backgrounds, limits are set on the production of dark matter particles. The observed limit on the cross section is dependent on the mass of the dark matter particle and the nature of its interactions with the SM particles. The upper limits on the dark matter production cross sections, as a function of $M_{\chi}$, for the vector and axial-vector interactions can be converted to lower limits on the effective contact interaction scale $\lambda$. These are then translated to an upper limit on the dark matter-nucleon scattering cross section, within the effective theory framework~\cite{bib:roniLHC}. Figure~\ref{fig:limits} shows the 90$\%$ CL upper limits on the dark matter-nucleon scattering cross section as a function of the mass of the dark matter particle for the spin dependent and spin independent interactions. Also shown are the results from other experiments, including the CDF monojet analysis~\cite{bib:CDFmonojet}. For spin dependent interactions, the bounds from the CMS monojet and monophoton analyses surpass all previous constraints for the 1$-$200 GeV mass range. For spin independent interactions, these limits extend the excluded $M_{\chi}$ range into the previously inaccessible region below 3.5 GeV.

In summary, results are presented for searches for dark matter at CMS using the monojet and monophoton plus \Met\ signatures. The observed event yields are consistent with those expected from SM backgrounds. Hence, limits are set on the dark matter-nucleon scattering cross section as a function of $M_{\chi}$ and represent a significant extension to those from previous experiments.


\section*{References}
\begin{thebibliography}{99}

\bibitem{bib:tait} M. Beltran {\it et al} \Journal{\JHEP}{09}{037}{2010}.
\bibitem{bib:tait2} J. Goodman {\it et al} \Journal{\PRL}{B695}{185}{2011}.
\bibitem{bib:tait3} J. Goodman {\it et al} \Journal{\PRD}{82}{116010}{2010}.
\bibitem{bib:tait4} A. Rajaraman {\it et al} arXiv:1108.1196.
\bibitem{bib:roniTevatron} Y. Bai {\it et al} \Journal{\JHEP}{12}{048}{2010}.
\bibitem{bib:roniLHC}P.J. Fox {\it et al}, \Journal{\PRD}{85}{056011}{2012}.
\bibitem{bib:particleflow}CMS Physics Analysis Summary CMS-PAS-PFT-09-001 (2009).
\bibitem{bib:antikt}M. Cacciari {\it et al} \Journal{JHEP}{04}{063}{2008}.
\bibitem{bib:monophoton} CMS Collaboration, arXiv:1204.0821.
\bibitem{bib:monojet} CMS Collaboration, \Journal{\PRL}{107}{201804}{2011}.
\bibitem{bib:CDFmonojet} CDF Collaboration, \Journal{\PRL}{108}{211804}{2012}.

\end{thebibliography}
\end{document}